\documentclass[prl,aps,twocolumn,showpacs,floatfix]{revtex4}

\usepackage{amsmath,amssymb}
\usepackage{graphicx}
\usepackage{version}
\DeclareGraphicsExtensions{.eps,.ps}

\newcommand{\bB}{{\bf B}}
\newcommand{\bQ}{{\bf Q}}
\newcommand{\bk}{{\bf k}}

\newcommand{\br}{{\bf r}}
\newcommand{\bp}{{\bf p}}
\newcommand{\bq}{{\bf q}}

\newcommand{\ba}{{\bf a}}

\newcommand{\beqa}{\begin{eqnarray}}
\newcommand{\eeqa}{\end{eqnarray}}

\begin{document}


\title
{Supersolid of indirect excitons in electron-hole quantum Hall systems}
\author{C.-H. Zhang}
\affiliation{Department of Physics, Indiana University-Purdue
University Indianapolis (IUPUI), Indianapolis, Indiana 46202, USA}
\author{Yogesh N. Joglekar}
\affiliation{Department of Physics, Indiana University-Purdue
University Indianapolis (IUPUI), Indianapolis, Indiana 46202, USA}
\date{\today}

\begin{abstract}
We investigate the ground state of a balanced electron-hole system
in the quantum Hall regime using mean-field theory and obtain a rich
phase diagram as a function of interlayer distance $d$ and the
filling factor within a layer. We identify an excitonic condensate
phase, an excitonic supersolid phase, as well as uncorrelated Wigner crystal
states. We find that balanced electron-hole system exhibits a
supersolid phase a wide range of filling factors, with different
crystal structure ground states. We obtain the ground state
stiffness in the excitonic phases and show that the phase
transitions from a uniform condensate to a supersolid is accompanied
by a marked change in the stiffness. Our results provide the first 
semi-quantitative determination excitonic supersolid phase diagram and 
properties.
\end{abstract}

\maketitle


\noindent{\it Introduction:} Over the past decade, Bose-Einstein
condensation of indirect excitons~\cite{Griffin95} in optically
pumped and doped electron-hole double quantum wells has been
extensively explored\ \cite{butov,snoke,lai,Butov04,Seamons06} The
concurrent theoretical analysis has largely focused on either a
uniform condensate, where a macroscopic number of excitons occupy
the zero-momentum state, or the condensation in a trap where the
translational symmetry is broken 
explicitly\ \cite{shev,Littlewood96,ga,palo,keeling}. On the other hand, recent
observations of the supersolid phase in $^4$He~\cite{Kim04,kim07} 
have revived the interest in and questions about
excitonic Bose-Einstein condensates with {\it spontaneously broken}
symmetries and their properties\ \cite{sasaki,Phillips07}.
(``Supersolid'' phase of cold atoms in optical lattices has been
extensively discussed, although, in that case, the translational
symmetry is explicitly broken by the optical lattice.)

Lozovik and Berman\ \cite{Lozovik98} first discussed the coherent
charge-density-wave (CCDW) ground state of indirect excitons in
electron-hole double quantum wells. Recently, based on general
principles, it was shown that electron-hole systems must support a
supersolid phase of excitons due to their dipolar repulsion\
\cite{Joglekar06}. Although a simple qualitative analysis implies
the existence of a supersolid phase in electron-hole bilayer
systems, the quantitative determination of the phase-boundary and
supersolid properties is made difficult by the dispersion of
electron (hole) bands.

\begin{figure}[thbf]
\includegraphics[width=1\columnwidth]{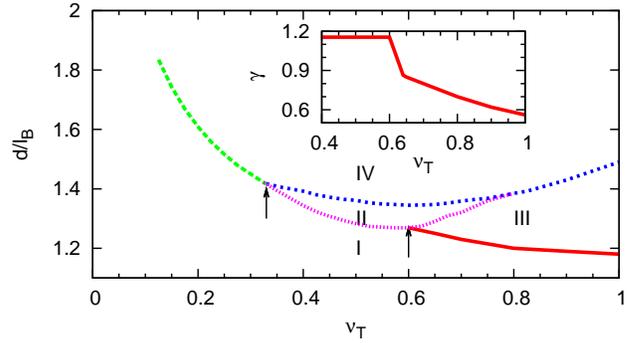}
\vspace{-8mm} \caption{(Color online) Ground state phase diagram of
electron-hole quantum Hall system. For small $d<d_{c_1}$ (region I),
the ground state is a uniform excitonic condensate; at large
$d>d_{c_2}$ (region IV), the ground state is uncorrelated triangular
Wigner crystals. For $d_{c_1}\leq d\leq d_{c_2}$ the ground state is a 
{\it supersolid of excitons}, with triangular (region II) and 
anisotropic (region III) lattice structures respectively. This state has
spontaneous interlayer phase coherence as well as spontaneously
broken translational symmetry. The inset shows that ground state
lattice anisotropy $\gamma(\nu_T)$ changes discontinuously at
the boundary between regions II and III.} \label{fig:dc1}
\end{figure}

In this paper, we study an electron-hole system in a strong magnetic
field, where the kinetic energy of carriers is quenched\ \cite{ja}. 
We consider a system with equal electron and hole filling factors
$\nu_e=\nu_h=\nu_T/2\leq 1/2$. CCDW ground states of such a system
have been investigated\ \cite{Chen91,Chen92}; however, such states
are destabilized by fluctuations\ \cite{Cote00,brey2000}. We obtain
the ground state phase diagram in the $(d,\nu_T)$ plane (Fig.\
\ref{fig:dc1}). Particle-hole symmetry in the lowest Landau level
maps the electron-hole system at $\nu_T$ onto an electron-hole
system at $2-\nu_T$ and implies that the phase diagram is symmetric 
around $\nu_T=1$. We
verify that this exact relation is satisfied by our mean-field
results. The phase diagram of a closely related
system - bilayer quantum Hall system where carriers in both layers
have the same polarity, near total filling factor $\nu_T=1$ - has
been extensively explored\ \cite{ja}. These systems have provided
signatures of excitonic condensation in interlayer tunneling and
counterflow experiments at small $d$~\cite{ja} and in the presence 
of a bias voltage~\cite{tutuc}.

For an electron-hole quantum Hall system, we find that (Fig.\
\ref{fig:dc1}): i) for small $d\leq d_{c_1}$ (region I), the ground
state is a uniform excitonic condensate irrespective of the total
filling factor ii) for large $d\geq
d_{c_2}$ (region IV), the ground state consists of uncorrelated
{\it triangular} Wigner crystals in the electron layer and the hole layer 
iii) for $d_{c_1}\leq d \leq d_{c_2}$, the ground state
is a {\it supersolid of excitons}, i.e. a state with spontaneous
interlayer phase coherence and spontaneously broken translational
symmetry. This supersolid has either a triangular lattice structure
(region II) or an anisotropic lattice structure (region III). To study the
robustness of phase coherence, we calculate the ground state
stiffness $\rho_s(d,\nu_T)$ by considering the mean-field energy
dependence on an in-plane magnetic field. We find that the phase
transitions with increasing $d$ - from a uniform excitonic
condensate to a supersolid to Wigner crystals - are accompanied by
marked changes in the stiffness.

In the following section, we briefly sketch the details of mean-field 
calculations. Then we discuss the phase diagram (Fig.~\ref{fig:dc1}) 
focusing on the excitonic supersolid phases, and the dependence of 
stiffness $\rho_s(d)$ on the interlayer distance. We conclude the 
section with a comment on electron-hole quantum Hall systems at total 
filling factors $\nu_T$ and $2-\nu_T$. In the last section, we summarize 
our results. 


\noindent{\it Microscopic Model:} Let us consider a bilayer system
with electrons as carriers in the top layer and holes in the bottom
layer, in a uniform magnetic field
$\bB=B_{\perp}\hat{z}+B_{\parallel}\hat{x}$. The magnetic field
normal to the layers $B_{\perp}$ quenches the kinetic energy of the
carriers, whereas the in-plane magnetic field $B_{||}$ allows us to
study mean-field states with a winding interlayer phase. We assume a
vanishing interlayer tunneling amplitude, keeping in mind that the
in-plane magnetic field cannot be gauged away for any nonzero
interlayer tunneling. The Hamiltonian for the system in the lowest
Landau level approximation is given by\ \cite{Chen91}
\begin{align}
\hat{H}=\frac{1}{2A}\sum_{\sigma_1\sigma_2,\bq}V_{\sigma_1\sigma_2}(\bq)
\hat{\rho}_{\sigma_1\sigma_1}(\bq)\hat{\rho}_{\sigma_2\sigma_2}(-\bq),
\end{align}
where $\sigma=e(h)$ denotes electron (hole), $A$ is the area of the
sample, $V_{ee}(\bq)=V_{hh}(\bq)= 2\pi e^2/(\epsilon q)$ is the
repulsive intralayer Coulomb interaction ($\epsilon\sim$ 10 is the
dielectric constant of the semiconductor) and
$V_{eh}(\bq)=-V_{ee}(\bq)\exp(-qd)$ is the interlayer attractive
Coulomb interaction. The momentum-space density operator in second
quantization is
\begin{align}
\hat{\rho}_{\sigma\sigma}(\bq)=\frac{1}{N_{\phi}}\sum_{k_1k_2}\langle
k_1| e^{-i\bq\cdot\br}|k_2\rangle c_{\sigma,k_1}^\dagger
c_{\sigma,k_2}
\end{align}
where $c_{\sigma,k}$ ($c^\dagger_{\sigma,k}$) is the annihilation
(creation) operator for a particle in the lowest Landau level state
$|k\rangle=|n=0,k\rangle$, $N_\phi=A/(2\pi l_B^2)$ is the degeneracy
of a single Landau level, and $l_B=\sqrt{hc/eB_{\perp}}$ is the
magnetic length. We define the excitonic condensate operator as
\begin{align}
\hat{\rho}_{eh}(\bq)&=\frac{1}{N_{\phi}} \sum_{k_1 k_2}
\langle k_1| e^{-i\bq\cdot\br}|k_2\rangle c^{\dagger}_{e,k_1}
c^{\dagger}_{h,-k_2}
\end{align}
and recall that single-particle wave-functions for holes are obtained by
complex-conjugation from the single-particle wavefunctions for electrons,
$\langle\br|n,k\rangle_h=\langle\br|n,k\rangle_e^{*}$. Following standard
procedure, we obtain the Hartree-Fock Hamiltonian
\begin{align}
\label{eq:hhf}
\hat{H}_{HF}&=\frac{N_{\phi}e^2}{\epsilon
l_B}\sum_{\sigma\bq}\left[U_{\sigma\sigma}(\bq)\hat{\rho}_{\sigma\sigma}(\bq)
+ U_{\sigma\bar{\sigma}}(\bq)\hat{\rho}_{\bar{\sigma}\sigma}(\bq)
\right].
\end{align}
The first term,
$U_{\sigma\sigma}(\bq)=\left[V_a(\bq)-V_b(\bq)\right]
\rho_{\sigma\sigma}(-\bq)-V_c(\bq)\rho_{\bar{\sigma}\bar{\sigma}}(-\bq)$
contains the intralayer Hartree and exchange, and interlayer Hartree
contributions respectively ($\bar{e}=h$), and second term
$U_{\sigma\bar{\sigma}}(\bq)=-V_d(\bq)
\rho_{\sigma\bar{\sigma}}(-\bq)$ denotes the excitonic condensate
contribution. These dimensionless contributions are given by
$V_a(\bq)=1/ql_B$, $V_b(\bq)=\sqrt{\pi/2}I_0(q^2l_B^2/4)$,
$V_c(\bq)=-e^{-qd}V_a(\bq)$, and
\begin{align}
V_d(\bq)=\int\frac{d^2p}{(2\pi)}l_B^2 V_a(\bp)
e^{-pd}e^{i\bp\times\bq\cdot\hat{z}l_B^2}.
\end{align}

To obtain the self-consistent density matrices, we introduce a
two-component operator $a^\dagger_{k}=[c^{\dagger}_{e,k},
c_{h,-k}]$ and define the $2\times2$ matrix Green's function~\cite{Fetter}
\begin{align}
G(\bQ;\tau)=-\frac{1}{N_{\phi}} \sum_{k_1k_2}\langle k_1|
e^{-i\bQ\cdot\br}|k_2\rangle\langle
\mbox{T}_{\tau}a_{k_1}(\tau)a^\dagger_{k_2}(0)\rangle.
\end{align}
The electron and hole density matrices $\langle\hat{\rho}_{\sigma\sigma}(\bQ)
\rangle$ as well as the (complex) excitonic order parameter
$\langle\hat{\rho}_{eh}(\bQ)\rangle$ are then determined from the equal-time
limit ($\tau\rightarrow 0^{-}$) of this Green's function matrix. In the
Hartree-Fock approximation, the equation of motion for $G$
matrix in the frequency space is given by\ \cite{Chen91,Cote91,Lian95}
\begin{align}
\label{eq:G_eq}
&\delta_{\bQ_i,0}I=\left[\begin{array}{cc}i\omega_n+\mu & 0
\\
0 & i\omega_n-\mu\end{array}\right]{\cal G}(\bQ_i;i\omega_n)
\nonumber\\
&-\sum_{j}\left[\begin{array}{cc} \Sigma_{ee}(\bQ_i,\bQ_j)
&\Sigma_{eh}(\bQ_i,\bQ^-_j)\\
\Sigma_{he}(\bQ^-_i,\bQ_j)
&-\Sigma_{hh}(\bQ^-_i,\bQ^-_j)\end{array}\right]{\cal G}(\bQ_j;i\omega_n),
\end{align}
where $I$ is the identity matrix, $\bQ^{\pm}=\bQ\pm\bk_B$,
$\bk_Bl_B^2=\hat{y}dB_{||}/B_{\perp}$ represents the relative displacement of
single-particle wavefunctions in the two layers due to the Lorenz
drift caused by the in-plane field~\cite{Lian95}, and we have defined a
modified Green's function
\begin{align}
\label{eq:G}
{\cal G}(\bQ)=\left[\begin{array}{cc}
G_{ee}(\bQ) & G_{eh}(\bQ^{+})e^{i\bk_B\times\bQ\cdot\hat{z}/2}\\
G_{he}(\bQ^-) & G_{hh}(\bQ)e^{i\bk_B\times\bQ\cdot\hat{z}/2}
\end{array}\right].
\end{align}
The self-energies in Eq.\ (\ref{eq:G_eq}) are given by
$\Sigma_{\sigma_1\sigma_2}(\bQ_i,\bQ_j)=U_{\sigma_1\sigma_2}(\bQ_{ij})
\exp(i\bQ_i\times\bQ_j\hat{z}l_B^2/2)$ with $\bQ_{ij}=\bQ_i-\bQ_j$.
The modified Green's function can be expressed as 
\begin{align}
\label{eq:calg} {\cal
G}(\bQ;i\omega_n)=\sum_{k}\frac{\lambda_{k}(\bQ)\lambda_k^{\dagger}(0)}
{i\omega_n-\omega_k},
\end{align}
where $\lambda_k^{\dagger}(\bQ)=[V_k^{*}(\bQ), U_k^{*}(\bQ^{-})]$
are the eigenvectors of the self-energy matrix with eigenvalue
$\omega_k$,
\begin{align}
\label{eq:diag} &\sum_{j}\left[\begin{array}{cc}
\Sigma_{ee}(\bQ_i,\bQ_j)-\mu\delta_{ij}
&\Sigma_{eh}(\bQ_i,\bQ^-_j)\\
\Sigma_{he}(\bQ^-_i,\bQ_j) &
\mu\delta_{ij}-\Sigma_{hh}(\bQ^-_i,\bQ^-_j)\end{array}\right]
\lambda_k(\bQ_j)\nonumber\\
&=\omega_k\lambda_k(\bQ_i)
\end{align}
We notice that, in contrast to a similar analysis for bilayer
quantum Hall systems\ \cite{Cote91,Lian95} - the chemical potential
appears with opposite signs in the electron and hole self-energies. 
The Hartree-Fock ground state energy  $E_{HF}(k_B)$ is calculated 
using the self-consistent density matrix using Eq.(\ref{eq:hhf}).

Equations\ (\ref{eq:G})-(\ref{eq:calg}) provide the set of equations
that are iteratively solved. We check that the resulting density matrix
satisfies the sum-rule~\cite{Cote91}
\begin{align}
\label{eq:sum}
\sum_{\bQ}\left[|\rho_{ee}(\bQ)|^2+|\rho_{eh}(\bQ)|^2\right]=
\rho_{ee}(0)=\nu_T/2.
\end{align}


\noindent {\it Results:} In order to determine the phase diagram in
the $d-\nu_T$ plane, we compare the energies of uniform excitonic
condensate ($\rho_{eh}\neq 0$, $\rho\propto\delta_{\bq,0}$), Wigner 
crystal in each layer ($\rho_{eh}=0$,$\rho_{ee}=\rho_{hh}\propto
\delta_{\bq,\bQ}$), and supersolid ($\rho_{eh}\neq 0$, 
$\rho_{ee}=\rho_{hh}\propto\delta_{\bq,\bQ}$) in the absence of an 
in-plane magnetic field $B_{||}=0$. We use simplified anisotropic 
lattice with two primitive lattice vectors\\cite{brey2000} 
$\ba_1=(a,b/2)$ and $\ba_2=(0,b)$, and define the lattice anisotropy 
$\gamma=b/a$. The lattice constant $a$ is determined by the constraint
that the unit cell contains one electron-hole pair, and the optimal
value of $\gamma$ is obtained by minimizing the mean-field energy.
Note that the triangular lattice ($\gamma=2/\sqrt{3}\approx$ 1.155) and
a stripe lattice ($\gamma\rightarrow 0$) are its special cases. (We
have found that square lattices have higher energy in the supersolid
and uncorrelated Wigner crystal phases.)

Figure\ \ref{fig:dc1} shows the ground state phase diagram.  At
small values of $d\leq d_{c_1}$, we find a uniform excitonic
condensate as the ground state over the entire range of total
filling factor (region I). For such a state, the mean-field
equations can be analytically solved and we obtain the layer
densities $\rho_{ee}(0)=\rho_{hh}(0)=\nu_T/2$ and the excitonic
condensate order parameter $\rho_{eh}(0)=\sqrt{\nu_T(2-\nu_T)}/2$.
These analytical results satisfy Eq.\ (\ref{eq:sum}) and imply that
the uniform phase coherence order parameter is the same for systems
with filling factors $\nu_T$ and $2-\nu_T$. The monotonic increase
in $d_{c_1}$ with decreasing $\nu_T$\ \cite{Chen91} is consistent
with observed strengthening of excitonic condensate phase in bilayer
quantum Hall systems with layer imbalance\ \cite{tutuc}. At large
values of $d\geq d_{c_2}$ (which is the same as $d_{c_1}$ for
$\nu_T\lesssim 1/3$), the ground state is uncorrelated triangular
Wigner crystals in the electron layer and the hole layer (region
IV).

\begin{figure}[h]
\includegraphics[width=1\columnwidth]{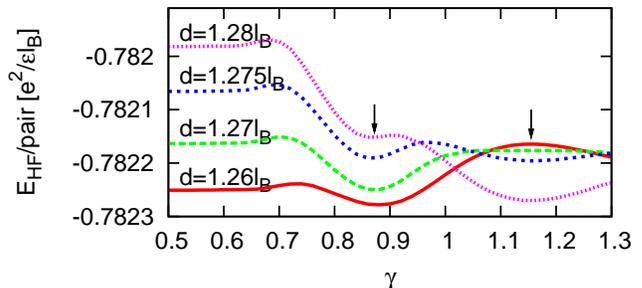}
\vspace{-8mm} \caption{(Color online) Mean-field energy as a
function of lattice anisotropy $\gamma$ for different values of
$d/l_B$ at total filling factor $\nu_T=0.62$. For $d/l_B=1.26$, the
optimal value of $\gamma\sim 0.89$ corresponds to an anisotropic
supersolid, whereas increasing $d/l_B$ shifts the optimal value to
$\gamma=2/\sqrt{3}=1.155$ corresponding to a triangular supersolid.}
\label{fig:nu0.62}
\end{figure}

For $\nu_T\ge 1/3$, we find that the ground state is a supersolid at
intermediate values of distance $d_{c_1}\leq d\leq d_{c_2}$. In
region II, the supersolid has a triangular lattice structure
($\gamma=2/\sqrt{3}$) whereas in region III, the lattice is
anisotropic. (The anisotropic supersolid with small $\gamma$ is similar 
to the CCDW solutions discussed in Ref.\cite{Chen91}). 
The boundary between two regions 
corresponds to a line of first-order phase transitions from a
triangular lattice to an anisotropic lattice. The inset in Fig.\
\ref{fig:dc1} shows the ground state lattice anisotropy
$\gamma(\nu_T)$ along the line $d_{c_1}$. For $\nu_T\leq 0.6$,
$\gamma=2/\sqrt{3}$ is constant, as expected for a triangular
lattice; near $\nu_T\approx 0.6$, $\gamma$ drops discontinuously and
then reduces monotonically with $\nu_T$. This discontinuity in
$\gamma$ indicates a first order transition from a triangular to an
anisotropic supersolid. The reverse transition, from region III to 
region II, can be induced by
increasing the interlayer distance $d$ at a given value of $\nu_T$. 
Figure\ \ref{fig:nu0.62} shows 
the mean-field energy as function of lattice anisotropy $\gamma$ for
increasing $d$. We see that the optimal value of $\gamma$ jumps 
discontinuously from $\gamma\approx 0.9$ (anisotropic lattice) to
$\gamma=2/\sqrt{3}$ (triangular lattice). The triangular lattice
structure in regions II and IV suggests that the phase transition
between them is a continuous phase transition. On the other hand,
different lattice structures in region III and IV imply that the
transition between them will be a first order transition.

\begin{figure}[h]
\includegraphics[width=1\columnwidth]{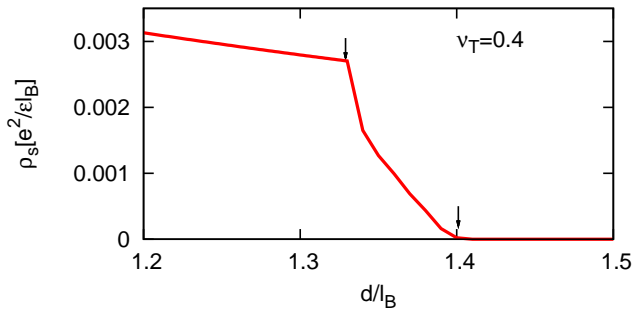}
\vspace{-8mm} \caption{(Color online) Dependence of ground-state
stiffness on interlayer distance $\rho_s(d)$ for total filling factors
$\nu_T=0.4$. The slope discontinuities in the stiffness occur at the
critical layer separations $d_{c_1}$ and $d_{c_2}$ respectively.}
\label{fig:rhos}
\end{figure}

An important property of a Bose-Einstein condensate, uniform or
otherwise, is the energy cost associated with spatial variations of
its phase. A uniform in-plane magnetic field induces such a
variation in a uniform excitonic condensate\ \cite{sasha}. We
calculate the stiffness, defined as the variation of ground state
energy with an in-plane magnetic field\ \cite{note}
\begin{align}
\label{eq:rhos} \rho_s=\frac{1}{A}\left.\frac{\partial^2
E_{HF}(k_B)}{\partial k_B^2}\right|_{k_B=0}
\end{align}
from the ground state energy. Figure\ \ref{fig:rhos} shows dependence of the
stiffness on interlayer distance $\rho_s(d)$ at $\nu_T=0.4$ We see a 
change in the stiffness slope at $d_{c_1}$, where the system 
changes from a uniform excitonic condensate to a triangular 
supersolid and $d_{c_2}$, where it changes from the supersolid to 
uncorrelated Wigner crystals, along with the expected monotonic decrease 
with $d$. Since the transport properties (in superfluid or supersolid phase) 
depend on the stiffness, generically, we expect that the
phase transitions will be visible in transport measurements.

\begin{figure}[htpb]
\includegraphics[width=1\columnwidth]{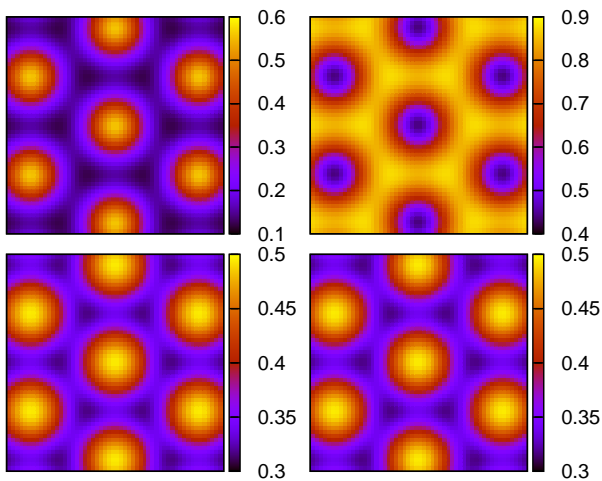}
\caption{(Color online) Ground state electron density
$\rho_{ee}(\br)$ (top) and excitonic order parameter
$|\rho_{eh}(\br)|$ (bottom), measured in units of $1/(2\pi l_B^2)$
at total filling factors $\nu_T=0.5$ (left) and $\nu_T=1.5$ (right).
The mean-field results satisfy
$\rho_{ee}(\br,\nu_T)=1-\rho_{ee}(\br,2-\nu_T)$ and $\rho_{eh}(\br,
\nu_T)=\rho_{eh}(\br,2-\nu_T)$.} \label{fig:ph}
\end{figure}

An electron-hole system at $\nu_T$ is mapped on to a system at $2-\nu_T$ 
after a particle-hole transformation on {\it both} layers. 
Figure\ \ref{fig:ph} shows that the mean-field density matrices, obtained 
separately for $\nu_T=0.5$ and $\nu_T=1.5$, indeed satisfy
$\rho_{ee}(\bq,2-\nu_T)=\delta_{\bq,0}-\rho_{ee}(\bq,\nu_T)$ and
$\rho_{eh}(\bq,2-\nu_T)=\rho_{eh}(\bq,\nu_T)$. Therefore, it is
enough to restrict ourselves to $\nu_T\leq 1$.


\noindent{\it Discussion:} The supersolid phase, with two
spontaneously broken continuous symmetries, has been a source of
extensive investigations in $^4$He\ \cite{Kim04,kim07,sasaki}, but has not
been experimentally realized in any other system. Our results
predict that electron-hole quantum Hall systems exhibit a rich phase
diagram including a supersolid phase robust over filling factors
$\nu_T\gtrsim1/3$. We show that the supersolid phase exhibits a
nonzero stiffness and obtain the dependence of stiffness $\rho_s(d)$
on interlayer distance. It is well known that mean-field approach
overestimates the stability ordered states. However, it is not clear
whether it overestimates the stability of one type of order
(excitonic condensation) over another (broken translational
symmetry); therefore, we believe that our conclusions - existence of
the supersolid phase in the intermediate distance regime and the
accompanying change in the phase stiffness - are generically valid. 

Our results present an example of a supersolid whose properties can be
investigated starting from a microscopic Hamiltonian; in particular, the
study of low-energy  excitations and transport properties in the supersolid
phase is possible. We point out that this excitonic system, although
analytically tractable, is significantly different from the zero-field system
in which, at low exciton densities, the excitons behave as non-interacting
bosons. Nonetheless, an experimental verification (or  falsification) of our
predictions will provide a better understanding of the supersolid  phase and
the properties of excitons in the quantum Hall regime.


\noindent{\it Acknowledgments:} It is a pleasure to thank Allan MacDonald 
for useful discussion.

\bibliographystyle{apsrev}
\bibliography{reference2}

\begin{thebibliography}{28}
\bibitem[{not()}]{note}
\bibinfo{note}{This stiffness, for a spatially homogeneous system, corresponds
  to the cost of phase fluctuations. Our calculations, with a supersolid ground
  state, include contributions from the phase variations and relative density
  displacements.}
\end{thebibliography}


\end{document}